%% file: main.tex
\documentclass[a4paper]{jpconf}
\usepackage{graphicx}
\usepackage{hyperref}

\begin{document}

\title{The Key4hep software stack: Beyond Future Higgs factories}
\author{Andre Sailer$^{1}$,
Benedikt Hegner$^{1}$,
Clement Helsens$^{2}$,
Erica Brondolin$^{1}$,
Frank-Dieter Gaede$^{3}$,
Gerardo Ganis$^{1}$,
Graeme A Stewart$^{1}$,
Jiaheng Zou$^{4}$,
Juraj Smiesko$^{1}$,
Placido Fernandez Declara$^{1}$,
Sang Hyun Ko$^{5}$,
Sylvester Joosten$^{6}$,
Tao Lin$^{4}$,
Teng Li$^{7}$,
Thomas Madlener$^{3}$,
Valentin Volkl$^{1}$,
Weidong Li$^{4}$,
Wenxing Fang$^{4}$,
Wouter Deconinck$^{8}$,
Xingtao Huang$^{7}$, and
Xiaomei Zhang$^{4}$
}

\address{$^{1}$CERN, Switzerland}
\address{$^{2}$KIT, Germany}
\address{$^{3}$Deutsches Elektronen-Synchrotron DESY, Germany}
\address{$^{4}$IHEP Beijing, China}
\address{$^{5}$Seoul National University, Korea}
\address{$^{6}$Argonne National Laboratory, Lemont, Illinois, USA}
\address{$^{7}$Shandong University, China}
\address{$^{8}$University of Manitoba, Winnipeg, Manitoba, Canada}

\ead{valentin.volkl@cern.ch, andre.philippe.sailer@cern.ch}

\begin{abstract}

  The Key4hep project aims to provide a turnkey software solution for the full experiment life-cycle, based on
  established community tools. Several future collider communities (CEPC, CLIC, EIC, FCC, and ILC) have joined to
  develop and adapt their workflows to use the common data model EDM4hep and common framework. Besides sharing of
  existing experiment workflows, one focus of the Key4hep project is the development and integration of new experiment
  independent software libraries. Ongoing collaborations with projects such as ACTS, CLUE, PandoraPFA and the
  OpenDataDector show the potential of Key4hep as an experiment-independent testbed and development platform. In this
  talk, we present the challenges of an experiment-independent framework along with the lessons learned from discussions
  of interested communities (such as LUXE) and recent adopters of Key4hep in order to discuss how Key4hep could be of
  interest to the wider HEP community while staying true to its goal of supporting future collider designs studies.

\end{abstract}

\input{introduction}
\input{challenges}

\section{Recent Developments in Key4hep}
\label{sec:recent}

Besides sharing of existing experiment workflows, one focus of the Key4hep project is the development and integration of
new experiment independent software libraries. Ongoing collaborations with projects such as ACTS~\cite{ai22:_common_track_softw_projec}, CLUE~\cite{Brondolin_2023,k4clue:gihub},
 PandoraPFA~\cite{Marshall:2015rfaPandoraSDK} and the OpenDataDector~\cite{corentin_allaire_2022_6445359,Gessinger-Befurt_2023} show the potential of Key4hep as an experiment-independent testbed and
development platform, and allow different communities to reap the benefits of new solutions, without the need for
different specific implementations. For the ACTS integration, recent developments to allow attaching ACTS specific
information to arbitrary DD4hep based geometries will streamline the integration of this track reconstruction framework.
The application of the CLUE calorimeter reconstruction algorithm for different high granularity calorimeters for CMS, or
CLIC like detectors also proves the effectiveness of providing generic solutions.

\begin{figure}[tbh]
\centering
\includegraphics[width=0.7\textwidth]{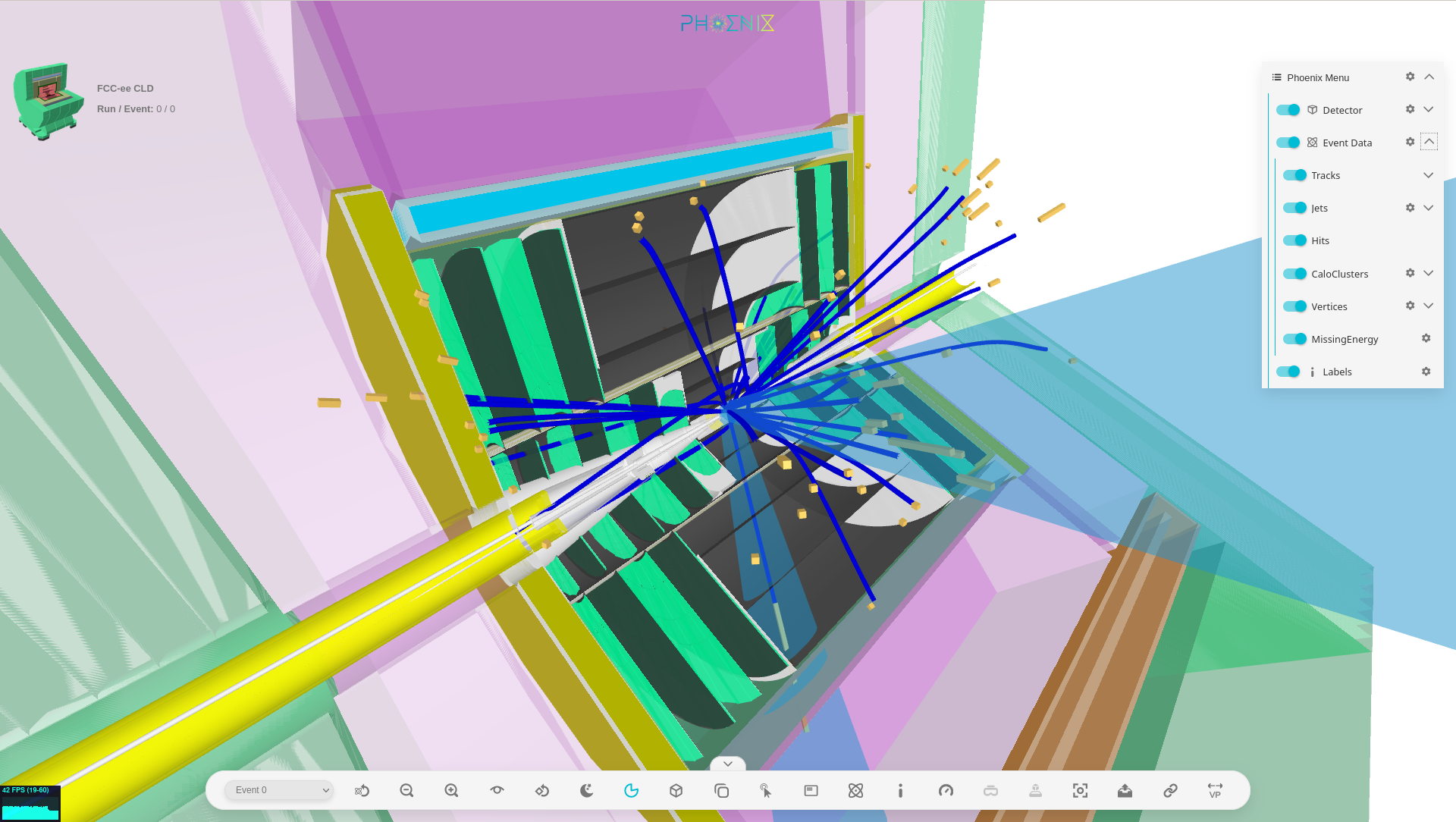}
\caption{\label{fig:phoenix}EDM4hep Event display using the Phoenix visualization tool.}
\end{figure}

Another example is the application of the Phoenix~\cite{fawad_ali_2023_7535649} event display. Fig.~\ref{fig:phoenix}
shows the event display for a DD4hep based detector using EDM4hep event data, converted to json via an EDM4hep utility, leveraging podio functionality.
A workflow that is viable for all experiments integrated in Key4hep.

\section{Key4hep and the Electron-Ion Collider}\label{sec:EIC}

\begin{figure}[tbh]
\centering
\includegraphics[width=0.7\textwidth]{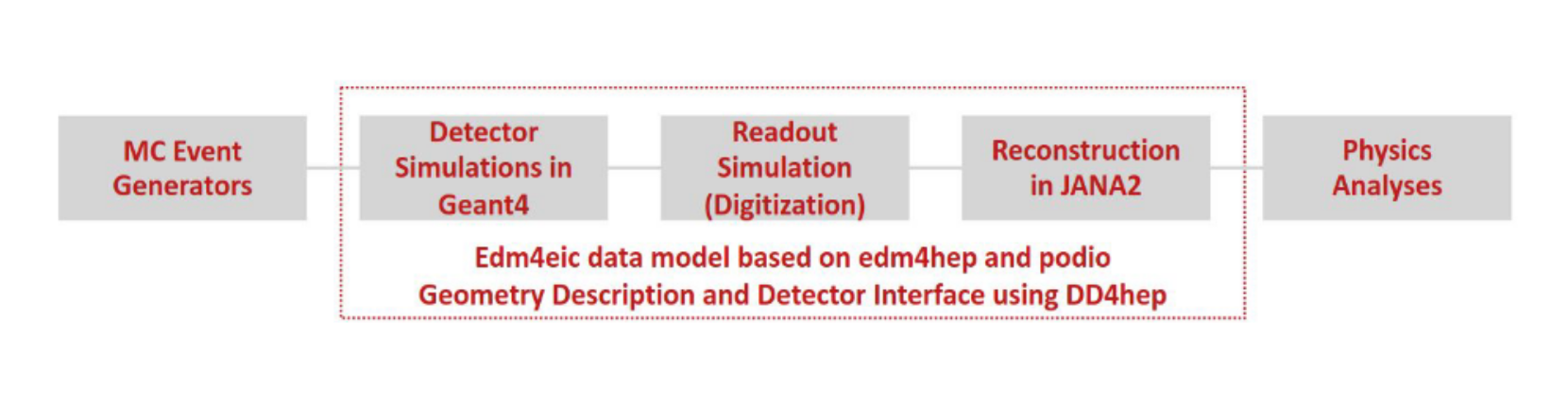}
\caption{\label{fig:eic}Diagram of steps and components of the Electron-Ion Collider software framework.}
\end{figure}

The Electron-Ion Collider (EIC) has chosen DD4hep for its geometry description, and uses DD4hep also for Geant4
simulations. It has also adopted podio to create an EDM4hep-based data model, adapted to its streaming readout, but uses
a different event processing framework: JANA2~\cite{JANA2} in place of Gaudi (Fig.~\ref{fig:eic}). The common use of EDM4hep allows a
broad collaboration nonetheless, and developments concerning a compatibility layer between Gaudi and JANA2 are ongoing.

\section{Key4hep and the Muon Collider}\label{sec:MuonColl}

\begin{figure}[tbh]
\centering
\includegraphics[width=0.7\textwidth]{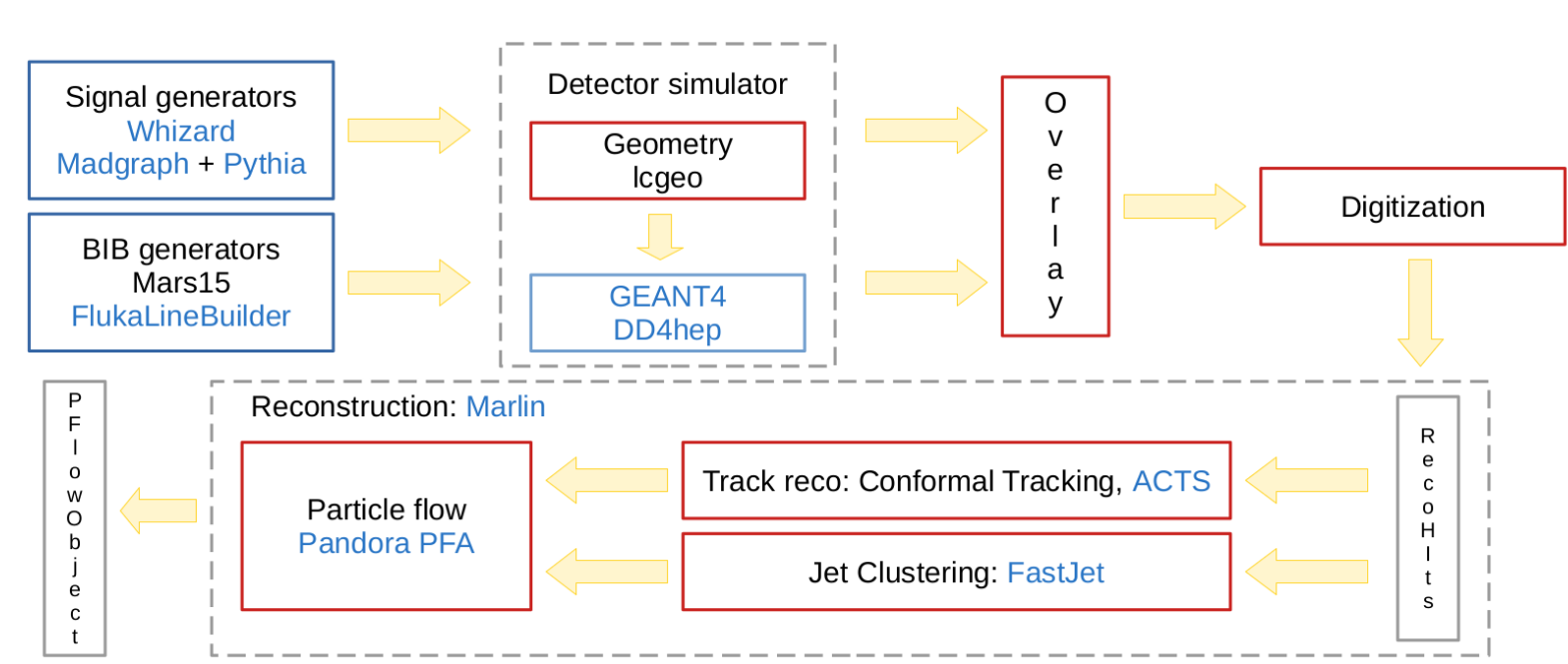}
\caption{\label{fig:mc}Diagram of steps and components of the Muon collider software framework.}
\end{figure}

The Muon Collider Design Study already uses large parts of iLCSoft for full simulations (Fig.~\ref{fig:mc}).
Discussion on full adoption of Key4hep are ongoing. Similar to the transition by the CLIC and ILC communities, the
k4MarlinWrapper compatibility layer can be used to seamlessly combine legacy iLCSoft components with new Key4hep
components~\cite{zenodo:k4marlinwrapper}. However, the large amount of background particles encountered in Muon
Collider collisions will put a higher memory pressure on the reconstruction workflow, which makes the in-memory
translation from EDM4hep to LCIO and vice-versa potentially too costly. Therefore, it will be necessary for some
algorithms to be ported from Marlin to Gaudi for this reason alone.

\input{luxe}

\section{Conclusion}

Beyond the existing collaboration members and interested future collider experiments, Key4hep can be adapted to the
needs of other planned and existing experiments. Especially small experiments may profit from a ready-to-use software
solutions, for example, LUXE. The shared development efforts across multiple smaller communities offers clear benefits,
despite possible challenges, that can be overcome through social or technical solutions.

\ack{This work benefited from support by the CERN Strategic R\&D Programme on
Technologies for Future Experiments (\url{https://cds.cern.ch/record/2649646/},
CERN-OPEN-2018-006).
This project has received funding from the European Union's Horizon
2020 Research and Innovation programme under grant agreement No. 101004761,
and it has received funding from the European Union's Horizon 2020 Research and Innovation programme under grant
agreement No. 871072.}

\section*{References}
\bibliographystyle{iopart-num.bst}
\bibliography{bibliography}

\end{document}

%% file: introduction.tex
\section{Introduction}\label{sec:introduction}

The Key4hep software framework offers a complete event processing framework for future collider experiments. It is based
on state-of-the art tools, such as Gaudi~\cite{Barrand:2001ny} as the event processing framework, DD4hep~\cite{dd4hep} for geometry
information, and podio~\cite{podio} to build its event data model EDM4hep~\cite{Gaede:2021izq}. Key4hep includes tools
from event generators, to full (Geant4~\cite{geant4} via DD4hep~\cite{ddsim}, k4SimGeant4~\cite{zenodo:k4simgeant4} and
Gaussino~\cite{lhcbdd4hep2019}) or fast simulation (Delphes~\cite{{deFavereau:2013fsa}}), to event reconstruction and
analysis. The CEPC, CLIC, FCC, and ILC communities for future Higgs factories are already in various stages of adoption
of this turnkey software stack~\cite{Fang:2023mwt,FernandezDeclara:2022voh}. In these proceedings we look at the current
status of Key4hep and also show that it can be a useful tool for experiments beyond future Higgs factories.


%% file: challenges.tex
\section{Challenges for an Experiment Independent Framework}
\label{sec:challenges}

One of the main challenges faced by an experiment independent, or inter-experiment, framework, is to ensure the
compatibility of developments with all participants. For example the event data model developments must not break
existing workflows, or only for very good reasons. Here the \emph{schema evolution} features of podio (see these
proceedings), will at least ensure backward compatibility with existing files. However, it is mandatory for all
communities to come to a consensus. The open developments of Key4hep on GitHub (\url{github.com/key4hep}) and the
regular meetings (\url{https://indico.cern.ch/category/11461/}) allow all voices to be heard.

Similarly, technical compatibility of new developments has to be ensured. The spack package
manager~\cite{Gamblin2015:spack}, used to build the Key4hep stack~\cite{Volkl:2021yvp}, allows us to build new
developments, including upstream and downstream packages, so that issues can be easily spotted. The use of the spack
package manager also lets everyone build the entire software stack by themselves, so that each community can deploy
software releases according to their own time constraints and requirements.

To ensure not only technical, but also physics performance, a validation framework for Key4hep is under development,
which will allow all communities to integrate their key performance indicators into a continuous validation system.


%% file: luxe.tex
\section{Usage of Key4hep in other experiments: the LUXE example}\label{sec:luxe}

Apart from the future collider projects there is also some interest to use
Key4hep from smaller experiments. One of these is the LUXE
experiment~\cite{LUXE_CDR}, planned to be built at DESY Hamburg aiming to study
non-perturbative QED effects in collisions of high-energy electrons or photons with a laser. From
a conceptual point of view, LUXE is somewhat comparable to a test beam target experiment with multiple
instruments, including spectrometers to analyse the residue of the collisions. The experiment will cover a wide range of
physical phase space, resulting in experimental conditions that will result in order of
magnitude differences in detector occupancies.

Currently discussions are ongoing on which parts of Key4hep can and should be
adapted by LUXE.\@ While several things are rather obvious choices, e.g., DD4hep
for detector description, or the general adaption of common workflows for
building software via spack, other components require more consideration. Among
these is the adaption of EDM4hep, which provides datatypes for describing, e.g.
the detector measurements from the positron tracking device. However, currently
it is completely lacking any representation of the measurements that would
result from, e.g., a 2D scintillator screen, imaged through a high resolution camera, that is used on the high rate side of the spectrometer,
or also measurements from a planned detector leveraging Cherenkov radiation.
Here some prototyping will have to be done before a decision can be made on
whether these datatypes could also be of general use in EDM4hep.
